\documentclass[12pt]{article}
 \usepackage{graphicx}
 \usepackage[cp1251]{inputenc} %% Windows
\begin{document}
 \noindent {\footnotesize\it Astronomy Letters, 2016, Vol. 42, No. 11, pp. 721--733.}
 \newcommand{\dif}{\textrm{d}}

 \noindent
 \begin{tabular}{llllllllllllllllllllllllllllllllllllllllllllll}
 & & & & & & & & & & & & & & & & & & & & & & & & & & & & & & & & & & & & & \\\hline\hline
 \end{tabular}

  \vskip 0.5cm
  \centerline{\large\bf Galactic Kinematics from Data on Open Star Clusters}
  \centerline{\large\bf from the MWSC Catalogue}
  \bigskip
  \bigskip
  \centerline{\large V.V. Bobylev$^1,$ A.T. Bajkova$^1$ and K.S. Shirokova$^{1,2}$ }
  \bigskip
  \centerline{\small\it $^1$ Pulkovo Astronomical Observatory, St. Petersburg,  Russia}
  \centerline{\small\it $^2$ St. Petersburg State University, St. Petersburg,  Russia}
  \bigskip
  \bigskip
{\bf Abstract}—Open star clusters from the MWSC (Milky Way Star
Clusters) catalogue have been used to determine the Galactic
rotation parameters. The circular rotation velocity of the solar
neighborhood around the Galactic center has been found from data
on more than 2000 clusters of various ages to be $V_0=236\pm6$ km
s$^{-1}$ for the adopted Galactocentric distance of the Sun
$R_0=8.3\pm0.2$ kpc. The derived angular velocity parameters are
 $\Omega_0=28.48\pm0.36$ km s$^{-1}$ kpc$^{-1}$,
 $\Omega'_0=-3.50\pm0.08$ km s$^{-1}$ kpc$^{-2}$, and
 $\Omega''_0= 0.331\pm0.037$ km s$^{-1}$ kpc$^{-3}$. The influence
of the spiral density wave has been detected only in the sample of
clusters younger than 50 Myr. For these clusters the amplitudes of
the tangential and radial velocity perturbations are
 $f_\theta=5.6\pm1.6$ km s$^{-1}$ and $f_R=7.7\pm1.4$ km s$^{-1}$, respectively; the
perturbation wavelengths are $\lambda_\theta=2.6\pm0.5$~kpc
 $(i_\theta=-11^\circ\pm2^\circ)$ and $\lambda_R=2.1\pm0.5$~kpc
 $(i_R=-9^\circ\pm2^\circ)$
for the adopted four-armed model $(m=4).$ The Sun's phase in the
spiral density wave is
 $(\chi_\odot)_\theta=-62^\circ\pm9^\circ$ and
 $(\chi_\odot)_R =-85^\circ\pm10^\circ$ from the residual
tangential and radial velocities, respectively.

%DOI: 10.1134/S1063773716100030

\section*{INTRODUCTION}
When the Galaxy and its subsystems are studied, great significance
is attached to the observational data quality. Open star clusters
(OSCs) play an important role here, because the mean values of
quite a few kinematic and photometric parameters derived from them
are highly accurate, while fitting to appropriate theoretical
isochrones based on probable cluster members allows one to
reliably estimate the distances (with an error of about 20\%) to
OSCs and their ages.

Open star clusters are used as a tool for studying the
interstellar dust (Trumpler 1930; Clayton and Fitzpatrick 1987;
Joshi 2005), the vertical structure of the Galactic disk (Bonatto
et al. 2006; Piskunov et al. 2006; Joshi 2007), its age (Twarog
and Anthony-Twarog 1989; Phelps 1997; Chaboyer et al. 1999), the
metallicity distribution in it (Netopil et al. 2016), the
age–metallicity relation (Friel 1995; Chen et al. 2003), and the
stellar evolution (Maeder and Mermilliod 1981; Moroni and
Straniero 2002). The great importance of OSCs for determining the
local parameters of the Galactic rotation curve (Glushkova et al.
1998; Zabolotskikh et al. 2002; Loktin and Beshenov 2003; Piskunov
et al. 2006; Bobylev et al. 2007) and the parameters of the spiral
density wave (Amaral and L\'epine 1997; Popova and Loktin 2005;
Loktin and Popova 2007; Naoz and Shaviv 2007; Bobylev et al. 2008;
L\'epine et al. 2008; Junqueira et al. 2015; Camargo et al. 2015)
has long been recognized.

The various data on OSCs are continuously updated. The number of
discovered clusters has increased considerably in recent years
owing to the appearance of large-scale infrared photometric
surveys. These include, for example, 2MASS (The Two Micron All Sky
Survey; Skrutskie et al. 2006) or WISE (Wide-field Infrared Survey
Explorer; Wright et al. 2010). The analysis of 2MASS data
performed by various authors enlarged the list of OSCs (Koposov et
al. 2005; 2008; Froebrich et al. 2007; Glushkova et al. 2010).
WISE data have revealed 652 new OSCs (Camargo et al. 2016) that
are virtually invisible in the optical band. The deep infrared VVV
(VISTA Variables in the Via Lactea; Cross et al. 2012) survey has
revealed 58 OSC candidates toward the central Galactic bulge
(Borisova et al. 2014), with the limiting distance being about
11.75 kpc (the cluster VVV CL117). Such works open up new
possibilities in studying the Galactic structure.

Infrared observations were also conducive to the appearance of
large-scale catalogues of stellar proper motions. These include,
for example, the XPM (Fedorov et al. 2009, 2011) or PPMXL (R\"oser
et al. 2010) catalogues. In the new MWSC (Milky Way Star Clusters;
Kharchenko et al. 2013) catalogue the mean proper motions were
determined for $\sim$3000 OSCs using data from the PPMXL
catalogue. It is important to note that the MWSC catalogue is
complete within 1.8 kpc of the Sun, which is more than twice the
completeness of the previous version of the catalogue compiled by
these authors. The goal of this paper is to refine the rotation
parameters of the Galaxy and its spiral structure using the latest
data on open star clusters. For this purpose we use the MWSC
catalogue.

 \section*{DATA}
The MWSC (Milky Way Star Clusters) catalogue was presented in
Kharchenko et al. (2013). It contains photometric and kinematic
data on 3006 Galactic objects (stellar associations, open and
globular clusters).

The MWSC catalogue surpasses considerably the previous COCD
catalogue of these authors (Kharchenko et al. 2005a, 2005b) in the
number of open star clusters with age and distance estimates and
kinematic data. To derive the mean proper motions of OSCs in MWSC,
we used the stellar proper motions from the PPMXL catalogue
(R\"oser et al. 2010), in which their random errors range from 4
to 10 mas yr$^{-1}$. In systematic terms, the PPMXL catalogue is
an extension of the International Celestial Reference System
(ICRS) to faint stars, because it is tied to the Hipparcos (1997)
catalogue.

The stellar proper motions in the PPMXL catalogue were obtained
from a completely different material than in the ASCC-2.5
catalogue (Kharchenko and R\"oser 2001), which served as a basis
for deriving the mean proper motions of OSCs in the COCD
catalogue. Therefore, with regard to the stellar proper motions,
we have a completely new material to investigate the kinematics of
OSCs. The CRVAD-2 compilation (Kharchenko et al. 2007) served to
derive the mean line-of-sight velocities of OSCs in the MWSC
catalogue, just as in the COCD catalogue. For some of the OSCs
information from other lists and databases, for example, from the
series of catalogues by Dias et al. (2002), was included in MWSC.

We supplemented the MWSC catalogue (Kharchenko et al. 2013) by new
data from two more papers of these authors: Schmeja et al. (2014)
on 139 high-latitude OSCs and Scholz et al. (2015) on 63 OSCs with
new distance and age determinations. We do not use the MWSC
objects with descriptors like ``a'', ``g'', ``n'', and ``s''
(associations, globular clusters, nebulae, and asterisms). As a
result, we produced a list of 2877 OSCs. For all these clusters
there are estimates of their ages, distances, and proper motion
components, while for 695 of them there are line-of-sight velocity
estimates.

 \section*{METHOD}
We know three stellar velocity components from observations: the
line-of-sight velocity $V_r$ and the two tangential velocity
components $V_l=4.74r\mu_l\cos b$ and $V_b=4.74r\mu_b$ along the
Galactic longitude $l$ and latitude $b$, respectively, expressed
in km s$^{-1}$. Here, the coefficient 4.74 is the quotient of the
number of kilometers in an astronomical unit by the number of
seconds in a tropical year, and $r$ is the heliocentric distance
of the star in kpc. The proper motion components $\mu_l\cos b$ and
$\mu_b$ are expressed in mas yr$^{-1}$. The velocities $U,V,W$
directed along the rectangular Galactic coordinate axes are
calculated via the components $V_r,V_l,V_b:$
 \begin{equation}
 \begin{array}{lll}
 U=V_r\cos l\cos b-V_l\sin l-V_b\cos l\sin b,\\
 V=V_r\sin l\cos b+V_l\cos l-V_b\sin l\sin b,\\
 W=V_r\sin b                +V_b\cos b,
 \label{UVW}
 \end{array}
 \end{equation}
where the velocity $U$ is directed from the Sun toward the
Galactic center, $V$ is in the direction of Galactic rotation, and
$W$ is directed to the north Galactic pole. We can find two
velocities, $V_R$ directed radially away from the Galactic center
and $V_{circ}$ orthogonal to it in the direction of Galactic
rotation, based on the following relations:
 \begin{equation}
 \begin{array}{lll}
  V_{circ}= U\sin \theta+(V_0+V)\cos \theta, \\
       V_R=-U\cos \theta+(V_0+V)\sin \theta,
 \label{VRVT}
 \end{array}
 \end{equation}
where the position angle $\theta$ obeys the relation
$\tan\theta=y/(R_0-x),$ and $x,y,z$ are the rectangular
heliocentric coordinates of the star (the velocities $U,V,W$ are
directed along the corresponding $x,y,z$ axes). The velocities
$V_R$ and $W$ barely depend on the pattern of the Galactic
rotation curve. However, to analyze the periodicities in the
tangential velocities, it is necessary to determine a smoothed
Galactic rotation curve and to form the residual velocities
$\Delta V_{circ}.$

To determine the parameters of the Galactic rotation curve, we use
the equations derived from Bottlinger's formulas, in which the
angular velocity $\Omega$ is expanded into a series to terms of
the second order of smallness in $r/R_0:$
\begin{equation}
 \begin{array}{lll}
 V_r=-U_\odot\cos b\cos l-V_\odot\cos b\sin l-W_\odot\sin b+\\
 +R_0(R-R_0)\sin l\cos b\Omega^\prime_0
 +0.5R_0(R-R_0)^2\sin l\cos b\Omega^{\prime\prime}_0,
 \label{EQ-1}
 \end{array}
 \end{equation}
 \begin{equation}
 \begin{array}{lll}
 V_l= U_\odot\sin l-V_\odot\cos l-r\Omega_0\cos b\\
 +(R-R_0)(R_0\cos l-r\cos b)\Omega^\prime_0
 +0.5(R-R_0)^2(R_0\cos l-r\cos b)\Omega^{\prime\prime}_0,
 \label{EQ-2}
 \end{array}
 \end{equation}
 \begin{equation}
 \begin{array}{lll}
 V_b=U_\odot\cos l\sin b + V_\odot\sin l \sin b-W_\odot\cos b-\\
 -R_0(R-R_0)\sin l\sin b\Omega^\prime_0
    -0.5R_0(R-R_0)^2\sin l\sin b\Omega^{\prime\prime}_0,
 \label{EQ-3}
 \end{array}
 \end{equation}
 where $R$ is the distance from the star to the Galactic rotation axis:
  \begin{equation}
 R^2=r^2\cos^2 b-2R_0 r\cos b\cos l+R^2_0.
 \end{equation}
The quantity $\Omega_0$ is the angular velocity of Galactic
rotation at the solar distance $R_0,$ the parameters $\Omega'_0$
and $\Omega''_0$ are the corresponding derivatives of the angular
velocity, and $V_0=|R_0\Omega_0|.$ As experience shows, to
construct a smooth Galactic rotation curve in the range of
distances $R$ from 2 to 12~kpc, it will suffice to know two
derivatives of the angular velocity, $\Omega'_0$ and $\Omega''_0$.
Note that the velocities $V_R$ and $\Delta V_{circ}$ must be freed
from the peculiar solar velocity $U_\odot,V_\odot,W_\odot.$

It is important to know the specific distance $R_0.$ Gillessen et
al. (2009) obtained one of its most reliable estimates,
$R_0=8.28\pm0.29$~kpc, by analyzing the orbits of stars moving
around the massive black hole at the Galactic center. From a
sample of masers with measured trigonometric parallaxes Reid et
al. (2014) estimated $R_0=8.34\pm0.16$~kpc; Bobylev and Bajkova
(2014a, 2014b) and Bajkova and Bobylev (2015) found
$R_0=8.3\pm0.4$~kpc also from masers. Based on these
determinations, we adopt $R_0=8.3\pm0.2$~kpc in this paper.

The influence of the spiral density wave in the radial, $V_R,$ and
residual tangential, $\Delta V_{circ},$ velocities is periodic
with an amplitude of $\sim$10 km s$^{-1}$. According to the linear
theory of density waves (Lin and Shu 1964), it is described by the
following relations:
 \begin{equation}
 \begin{array}{lll}
       V_R =-f_R \cos \chi,\\
 \Delta V_{circ}= f_\theta \sin\chi,
 \label{DelVRot}
 \end{array}
 \end{equation}
where
 \begin{equation}
 \chi=m[\cot(i)\ln(R/R_0)-\theta]+\chi_\odot
 \end{equation}
is the phase of the spiral density wave ($m$ is the number of
spiral arms, $i$ is the pitch angle of the spiral pattern, and
$\chi_\odot$ is the Sun's radial phase in the spiral density
wave); $f_R$ and $f_\theta$ are the amplitudes of the radial and
tangential velocity perturbations, which are assumed to be
positive.

In the next step, we apply a spectral analysis to study the
periodicities in the velocities $V_R$ and $\Delta V_{circ}.$ The
wavelength $\lambda$ (the distance between adjacent spiral arm
segments measured along the radial direction) is calculated from
the relation
\begin{equation}
 \frac{2\pi R_0}{\lambda}=m\cot(i).
 \label{a-04}
\end{equation}
Let there be a series of measured velocities $V_{R_n}$ (these can
be both radial, $V_R,$ and tangential, $\Delta V_{circ},$
velocities), $n=1,\dots,N,$ where $N$ is the number of objects.
The objective of our spectral analysis is to extract a periodicity
from the data series in accordance with the adopted model
describing a spiral density wave with parameters $f_R, f_\theta,
\lambda (i)$ and $\chi_\odot$.

Having taken into account the logarithmic behavior of the spiral
density wave and the position angles of the objects $\theta_n,$
our spectral (periodogram) analysis of the series of velocity
perturbations is reduced to calculating the square of the
amplitude (power spectrum) of the standard Fourier transform
(Bajkova and Bobylev 2012):
\begin{equation}
 \bar{V}_{\lambda_k} = \frac{1} {N}\sum_{n=1}^{N} V^{'}_n(R^{'}_n)
 \exp\Bigl(-j\frac {2\pi R^{'}_n}{\lambda_k}\Bigr),
 \label{29}
\end{equation}
where $\bar{V}_{\lambda_k}$ is the $k$th harmonic of the Fourier
transform with wavelength $\lambda_k=D/k,$ $D$ is the period of
the series being analyzed,
 \begin{equation}
 \begin{array}{lll}
 R^{'}_{n}=R_{\circ}\ln(R_n/R_{\circ}),\\
 V^{'}_n(R^{'}_n)=V_n(R^{'}_n)\times\exp(jm\theta_n).
 \label{21}
 \end{array}
\end{equation}
The algorithm of searching for periodicities modified to properly
determine not only the wavelength but also the amplitude of the
perturbations is described in detail in Bajkova and Bobylev
(2012).

The sought-for wavelength $\lambda$ corresponds to the peak value
of the power spectrum $S_{peak}.$ The pitch angle of the spiral
density wave is derived from Eq.~(9). We determine the
perturbation amplitude and phase by fitting the harmonic with the
wavelength found to the observational data. The following relation
can also be used to estimate the perturbation amplitude:
 \begin{equation}
f_R(f_\theta)=\sqrt{4\times S_{peak}}.
 \label{Speak}
 \end{equation}
Thus, our approach consists of two steps: (i) the construction of
a smooth Galactic rotation curve and (ii) a spectral analysis of
both radial, $V_R,$ and residual tangential, $\Delta V_{circ},$
velocities. Such a method was applied by Bobylev et al. (2008) to
study the kinematics of young Galactic objects, by Bobylev and
Bajkova (2012) to analyze Cepheids, and by Bobylev and Bajkova
(2013, 2015a) to determine the Galactic rotation curve from
massive OB stars.

 \section*{RESULTS}
The system of conditional equations (3)--(5) is solved by the
least-squares method with weights of the form
 $w_r=S_0/\sqrt {S_0^2+\sigma^2_{V_r}},$
 $w_l=S_0/\sqrt {S_0^2+\sigma^2_{V_l}}$ and
 $w_b=S_0/\sqrt {S_0^2+\sigma^2_{V_b}},$
where $S_0$ is the ``cosmic'' dispersion, $\sigma_{V_r},
\sigma_{V_l}, \sigma_{V_b}$ are the dispersions of the
corresponding observed velocities. $S_0$ is comparable to the
root-mean-square residual $\sigma_0$ (the error per unit weight)
in solving the conditional equations (3)--(5); therefore, we
adopted $S_0=15$~km s$^{-1}.$ The OSC distance errors were assumed
to be 20\%.

%%%%%%%%%%%%%%% Table~1.
\begin{table}[t]\caption[]{\small%\baselineskip=1.0ex
  Kinematic parameters found from OSCs by method I for four age intervals
  }
\begin{center}      \label{t1}
\begin{tabular}{|l|r|r|r|r|r|}\hline
 Parameters                   &    $\lg t<7.7$ &  $\lg t:7.7-8.4$ &  $\lg t:8.4-8.8$ &  $\lg t>8.8$ \\\hline
 $U_\odot,$  km s$^{-1}$      & $ 9.74\pm1.13$ & $ 8.38\pm1.01$ & $10.06\pm1.46$ & $18.40\pm2.00$ \\
 $V_\odot,$  km s$^{-1}$      & $11.19\pm1.38$ & $13.48\pm1.13$ & $10.94\pm1.57$ & $15.43\pm2.44$ \\
 $W_\odot,$  km s$^{-1}$      & $ 6.19\pm1.10$ & $ 5.97\pm1.02$ & $ 6.91\pm1.42$ & $ 7.78\pm2.14$ \\
     $\Omega_0,$ km s$^{-1}$ kpc$^{-1}$  & $28.60\pm0.81$ & $26.34\pm0.87$ & $24.64\pm1.56$ & $28.22\pm1.54$ \\
 $\Omega^{'}_0,$ km s$^{-1}$ kpc$^{-2}$  & $-4.04\pm0.16$ & $-3.51\pm0.17$ & $-3.61\pm0.29$ & $-3.29\pm0.34$ \\
$\Omega^{''}_0,$ km s$^{-1}$ kpc$^{-3}$  & $ 0.19\pm0.13$ & $ 0.40\pm0.15$ & $ 0.78\pm0.17$ & $ 0.20\pm0.17$ \\
   $\sigma_0,$   km s$^{-1}$  &          15.7  &          13.7  &          15.9  &          21.0  \\
         $N_\star$            &           209  &           197  &           136  &           123  \\
         $N_{equation}$       &           600  &           563  &           386  &           316  \\
             $A,$ km s$^{-1}$ kpc$^{-1}$  & $-16.66\pm0.64$ & $-14.57\pm0.70$ & $-15.00\pm1.18$ & $-13.64\pm1.42$ \\
             $B,$ km s$^{-1}$ kpc$^{-1}$  & $ 11.95\pm1.03$ & $ 11.78\pm1.11$ & $  9.65\pm1.96$ & $ 14.59\pm2.09$ \\
  \hline
\end{tabular}
\end{center}
\end{table}
%%%%%%%%%%%%%%%%%%%%%%%%%%%%%%% t-1...
%%%%%%%%%%%%%    T~2.
\begin{table}[t]\caption[]{\small%\baselineskip=1.0ex
  Kinematic parameters found from OSCs by method II for four age intervals
  }
\begin{center}      \label{t2}
\begin{tabular}{|l|r|r|r|r|r|}\hline
 Parameters                 &    $\lg t<7.7$ &  $\lg t:7.7-8.4$ &  $\lg t:8.4-8.8$ &  $\lg t>8.8$ \\\hline
 $U_\odot,$    km s$^{-1}$  & $11.42\pm1.14$ & $ 9.49\pm0.97$ & $11.74\pm1.15$ & $13.94\pm1.09$ \\
 $V_\odot,$    km s$^{-1}$  & $13.77\pm1.50$ & $13.64\pm1.24$ & $17.85\pm1.33$ & $21.96\pm1.29$ \\
 $W_\odot,$    km s$^{-1}$  & $ 6.53\pm1.01$ & $ 6.42\pm0.85$ & $ 7.19\pm0.87$ & $ 7.59\pm0.87$ \\
     $\Omega_0,$ km s$^{-1}$ kpc$^{-1}$ & $28.71\pm0.70$ & $26.71\pm0.72$ & $28.49\pm0.79$ & $28.76\pm0.64$ \\
 $\Omega^{'}_0,$ km s$^{-1}$ kpc$^{-2}$ & $-3.75\pm0.14$ & $-3.38\pm0.15$ & $-3.92\pm0.18$ & $-3.33\pm0.14$ \\
$\Omega^{''}_0,$ km s$^{-1}$ kpc$^{-3}$ & $ 0.20\pm0.11$ & $ 0.17\pm0.11$ & $ 0.82\pm0.09$ & $ 0.27\pm0.06$ \\
   $\sigma_0,$   km s$^{-1}$            &          20.2  &          18.2  &          21.0  &          26.0  \\
         $N_\star$            &           476  &           509  &           671  &          1221  \\
         $N_{equation}$       &          1016  &          1120  &          1338  &          2037  \\
  $A,$ km s$^{-1}$ kpc$^{-1}$ & $-15.54\pm0.60$ & $-14.02\pm0.64$ & $-16.27\pm0.73$ & $-13.82\pm0.56$ \\
  $B,$ km s$^{-1}$ kpc$^{-1}$ & $ 13.17\pm0.92$ & $ 12.69\pm0.97$ & $ 12.22\pm1.08$ & $ 14.94\pm0.85$ \\
  \hline
\end{tabular}
\end{center}
\end{table}
%%%%%%%%%%%%%%%%%%%%%%%%%%%%%%%

 \subsection*{Method I}
First we obtained a solution based on a sample of OSCs for which
the space velocities $U,V,W$ could be calculated. As can be seen
from Eqs. (1), such clusters must be provided with the proper
motions, line-of-sight velocities, and distances. In this case,
each OSC gives the three conditional equations (3), (4), and (5).
There are a total of 695 such clusters in the MWSC catalogue. We
selected the OSCs satisfying the following constraint on the
magnitude of the total space velocity:
 \begin{equation}
 \begin{array}{lll}
  \sqrt{U^2+V^2+W^2}<150~\hbox {km s$^{-1}$}.
 \label{criteri-1}
 \end{array}
\end{equation}
The $3\sigma$ criterion was applied in solving the conditional
equations (3)--(5). Since the proper motion errors increase
dramatically with distance, we restricted the use of clusters with
proper motions to the radius $r=4$~kpc. The latter restriction
cuts off only about 20 distant OSCs, while reducing considerably
(by 2--3\%) $\sigma_0.$ Based on our sample of 665 clusters, we
found the following kinematic parameters by this method:
 \begin{equation}
 \label{solution-1}
 \begin{array}{lll}
 (U_\odot,V_\odot,W_\odot)=(10.61,12.59,6.72)\pm(0.66,0.75,0.68)~\hbox{km s$^{-1}$},\\
      \Omega_0 =~27.14\pm0.53~\hbox{km s$^{-1}$ kpc$^{-1}$},\\
  \Omega^{'}_0 =-3.69\pm0.10~\hbox{km s$^{-1}$ kpc$^{-2}$},\\
 \Omega^{''}_0 =0.375\pm0.067~\hbox{km s$^{-1}$ kpc$^{-3}$}.
 \end{array}
 \end{equation}
In this solution the error per unit weight is $\sigma_0=16.5$ km
s$^{-1}$. For the adopted $R_0=8.3\pm0.2$ kpc the linear Galactic
rotation velocity $(V_0=|R_0\Omega_0|)$ is $V_0=225\pm7$ km
s$^{-1}$, while the Oort constants
 $(A=0.5\Omega'_0 R_0$ and $B=\Omega_0+0.5\Omega'_0 R_0)$ are
 $A=-15.31\pm0.43$ km s$^{-1}$ kpc$^{-1}$ and
 $B= 11.83\pm0.68$ km s$^{-1}$ kpc$^{-1}$

Table 1 gives four solutions obtained in this approach for four
age intervals. Apart from the values found for the six sought-for
unknowns in Eqs. (3)--(5), Table 1 gives $\sigma_0,$ the number of
OSCs $N_\star,$ the number of equations in system $N_{equation}$
(3)--(5) after all rejections, and the Oort constants $A$ and $B.$

 \subsection*{Method II}
In this approach we exploit all potentialities of the available
data. The clusters with the proper motions, line-of-sight
velocities, and distances give all three equations (3)--(5), while
the clusters for which only the proper motions are available give
only two equations, (4) and (5). Just as in the first method, we
restricted the use of clusters with known proper motions to the
radius $r=4$~kpc. More than 2000 clusters are involved in the
solution, while the total number of equations is
$N_{equation}=5496.$ The following kinematic parameters were found
in this approach:
 \begin{equation}
 \label{solution-2}
 \begin{array}{lll}
 (U_\odot,V_\odot,W_\odot)=(12.14,17.29,7.18)\pm(0.55,0.67,0.46)~\hbox{km s$^{-1}$},\\
      \Omega_0 =~28.48\pm0.36~\hbox{km s$^{-1}$ kpc$^{-1}$},\\
  \Omega^{'}_0 =-3.50\pm0.08~\hbox{km s$^{-1}$ kpc$^{-2}$},\\
 \Omega^{''}_0 =0.331\pm0.037~\hbox{km s$^{-1}$ kpc$^{-3}$}.
 \end{array}
 \end{equation}
In this solution the error per unit weight is $\sigma_0=22.3$ km
s$^{-1}$. The Galactic rotation velocity is $V_0=236\pm6$ km
s$^{-1}$ (for $R_0=8.3\pm0.2$~kpc), while the Oort constants
$A=-14.52\pm0.32$ km s$^{-1}$ kpc$^{-1}$ and $B=13.95\pm0.48$ km
s$^{-1}$ kpc$^{-1}$ show that the Galactic rotation curve in the
solar neighborhood is flatter than that in the case of solution
(14). Table 2 gives four solutions obtained in this approach for
four age intervals.

In comparison with solution (14), solution (15) has a higher value
of $\sigma_0.$ This is due to a considerable increase in the
random errors of the velocities $V_l$ and $V_b$ dependent on the
proper motion errors and the errors in the distance estimates.
Nevertheless, solution (15) has the following advantages over
solution (14): (i) the velocities $U_\odot$ and $V_\odot$ increase
with OSC age more gradually, (ii) a more stable angular velocity
$\Omega_0$ for various age intervals is observed, and (iii) the
errors in all six sought-for parameters decreased by a factor of
1.5.

The distribution of OSCs divided into four age groups on the
Galactic $XY$ plane is presented in Fig.~1. To construct this
figure, we used all 2877 OSCs without any restrictions. As can be
seen from the figure, only the youngest clusters tend to
concentrate toward the segments of the Carina–Sagittarius (number
II in Fig.~1) and Perseus (number III in Fig. 1) spiral arms and
the Local Arm (Bobylev and Bajkova 2014d).

 \subsection*{Perturbations in the Velocities from the Density Wave}
In both Table 1 and Table 2 the errors per unit weight $\sigma_0$
for the youngest clusters are larger than those for older
clusters. This is due to the influence of the spiral density wave
on the motion of the youngest clusters.

Note that the velocity perturbations from the density wave can be
analyzed by applying a spectral analysis only based on OSCs with
known space velocities. Therefore, to solve this problem, we took
209 youngest clusters $(\log t<7.7)$ and found the parameters
specified in the first column of Table 1 from them. The Galactic
rotation curve for these OSCs and their residual tangential,
$\Delta V_{circ},$ and radial, $V_R,$ velocities as a function of
distance $R$ are presented in Fig. 2, while the power spectra
corresponding to these velocities are presented in Fig. 3. The
rotation curve in Fig. 2a was constructed with the parameters from
the first column of Table 2, while the error in $R_0$ of 0.2 kpc
was taken into account when constructing the confidence region.

When constructing Fig. 2 and applying a spectral analysis using
Eqs. (7)--(11), we assumed the spiral pattern to be a four-armed
one $(m=4)$ with the pitch angle $i=-13^\circ.$ These parameters
are close to the present-day estimates of the geometrical
parameters of the Galactic spiral pattern (Bobylev and Bajkova
2014c; Hou and Han 2014; Valle\'e 2015). The signal amplitude is
easily estimated from the power spectrum using Eq. (12). We found
(see Fig. 3) that the perturbations in the radial velocities,
where $f_R=7.7\pm1.4$ km s$^{-1}$, manifest themselves with
increasing amplitude, while the amplitude of the periodicity in
the tangential velocities is only $f_\theta=5.6\pm1.6$ km
s$^{-1}$. The significance of the main peak is $p=0.983$ in Fig.
3b and only $p=0.884$ in Fig. 3a.

However, the perturbations in the residual tangential velocities
have a longer wavelength, $\lambda_\theta=2.6\pm0.5$ kpc. Based on
Eq. (9), we then find $i_\theta=-11^\circ\pm2^\circ$ (for
$R_0=8.3$ kpc), which agrees satisfactorily with $i=-13^\circ$
that we found from masers wit measured trigonometric parallaxes
(Bobylev and Bajkova 2014c). From the radial velocities we have
$\lambda_R=2.1\pm0.5$ kpc and, consequently,
$i_R=-9^\circ\pm2^\circ.$ As our analysis of various samples of OB
stars showed (Bobylev and Bajkova 2015a), the amplitude of the
perturbations in the radial velocities usually exceeds that in the
tangential ones. This is most likely due to the presence of
significant noise in the cluster velocities. The revealed waves
are shown in Figs. 2b and 2c. The Sun's phase in the spiral
density wave is seen to be $(\chi_\odot)_R=-85^\circ\pm10^\circ$
if this angle (increases toward the Galactic center) is measured
from the Carina--Sagittarius arm ($R\approx7.5$~kpc). The wave in
the tangential velocities has a shift by a value close to $\pi/2.$
The Sun's phase for this wave is
$(\chi_\odot)_\theta=-62^\circ\pm9^\circ.$ A peculiarity of our
approach (Eqs. (10)--(11)) is that we take into account the
logarithmic behavior of the spiral density wave. It can be seen
from Figs. 2b and 2c that the revealed waves have a distinct
logarithmic behavior (the wavelength $\lambda$ increases with
$R$).

%%%%%%%%%%%%%%%%%%%%%%%% FIG.1:
\begin{figure}[p]
{\begin{center}
   \includegraphics[width=0.9\textwidth]{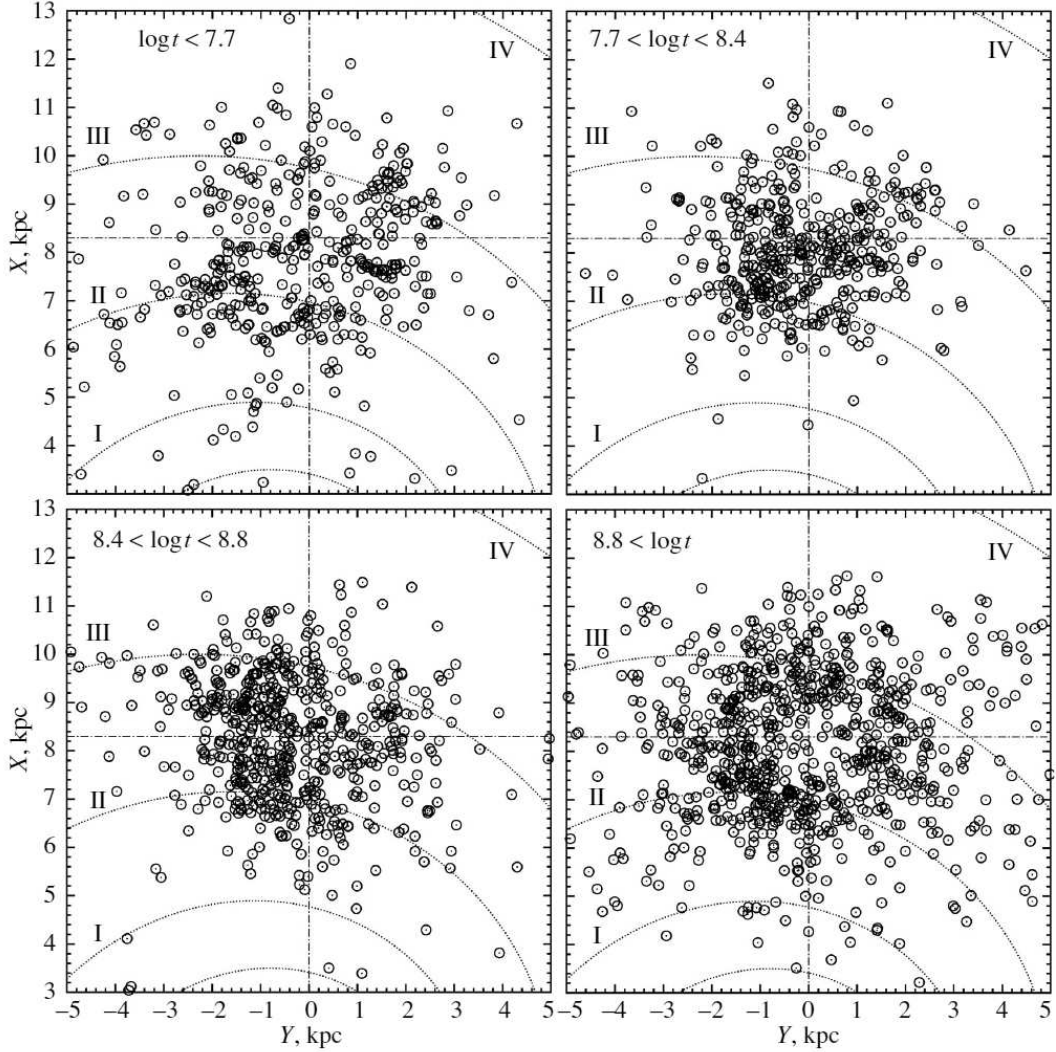}
 \caption{
 Distribution of four samples of OSCs with different ages on the Galactic $XY$ plane.
 The Sun's coordinates are
$(X,Y)=(8.3,0)$~kpc. The four-armed spiral pattern with the pitch
angle $i=-13^\circ$ found from masers (Bobylev and Bajkova 2014c)
is shown. The spiral arm segments are numbers by Roman numerals.
  } \label{f1}
\end{center}}
\end{figure}
%%%%%%%%%%%%%%%%%%%%%%%%%%%%%%%%%%%%%%%%%%%%%%%%%%%%%%%%%%%%%%%%%%%
%%%%%%%%%%%%%%%%%%%%%%%%%%%%%%%%%%%%%%%%%%%%%%%%%%%%%%%%%%%%%%%%%%%
 \begin{figure}
 {\begin{center}
 \includegraphics[width=0.6\textwidth]{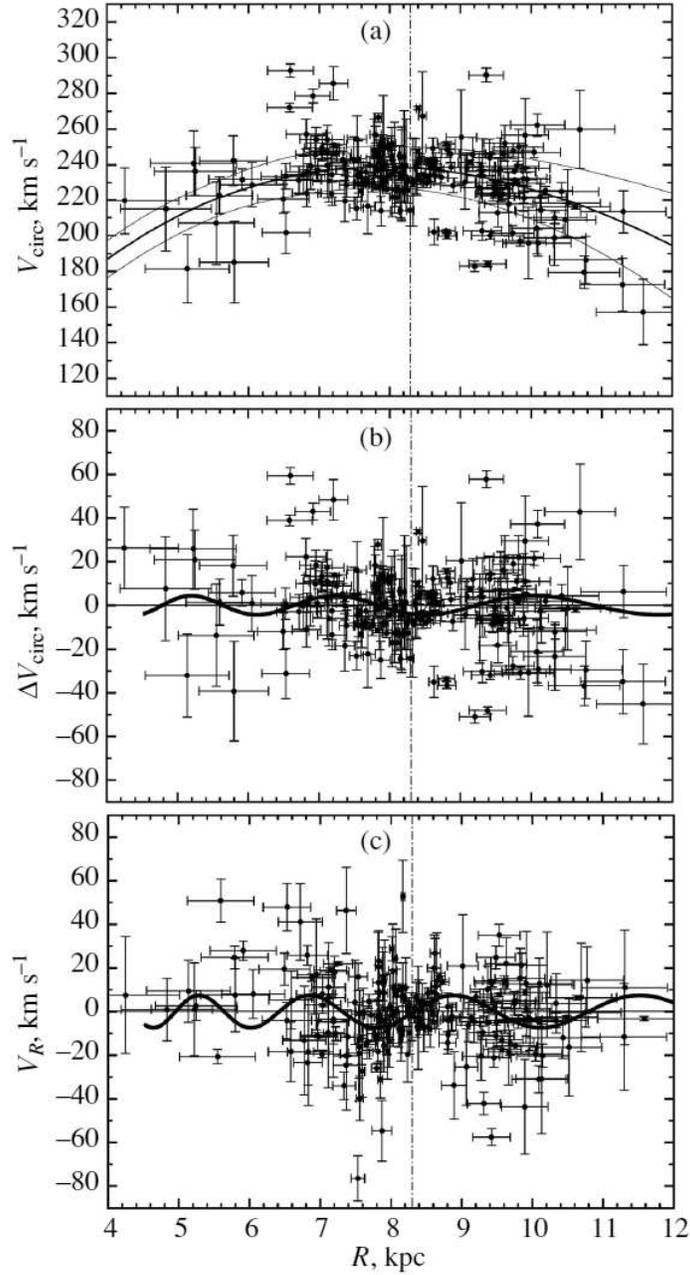}
 \caption{
(a) Galactic rotation curve constructed from the sample of 209
youngest OSCs $(\log t<7.7)$ with an indication of the boundaries
of the $1\sigma$ confidence intervals, (b) the residual rotation
velocities $\Delta V_{circ}$ of stars, and (c) the radial
velocities $V_R$ of stars; the vertical dash–dotted line marks the
Sun's position.
  }
  \label{f2}
 \end{center} }
 \end{figure}
%%%%%%%%%%%%%%%%%%%%%%%%%%%%%%%%%%%%%%%%%%%%%%%%%%%%%%%%%%%%%%%%%%%
%%%%%%%%%%%%%%%%%%%%%%%%%%%%%%%%%%%%%%%%%%%%%%%%%%%%%%%%%%%%%%%%%%%
 \begin{figure}
 {\begin{center}
 \includegraphics[width=0.5\textwidth]{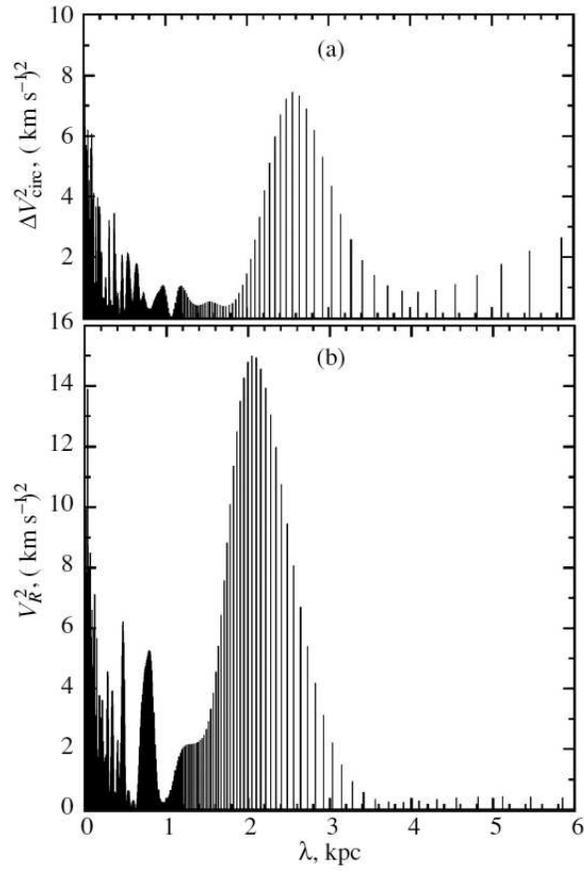}
 \caption{
Power spectrum of the residual tangential, $\Delta V_{circ}$ (a),
and radial, $V_R$ (b), velocities for the sample of 209 youngest
 $(\log t<7.7)$ OSCs.
 }
  \label{f3}
 \end{center} }
 \end{figure}
%%%%%%%%%%%%%%%%%%%%%%%%%%%%%%%%%%%%%%%%%%%%%%%%%%%%%%%%%%%%%%%%%%%
%%%%%%%%%%%%%%%%%%%%%%%%%%%%%%%%%%%%%%%%%%%%%%%%%%%%%%%%%%%%%%%%%%%
 \begin{figure}
 {\begin{center}
 \includegraphics[width=0.5\textwidth]{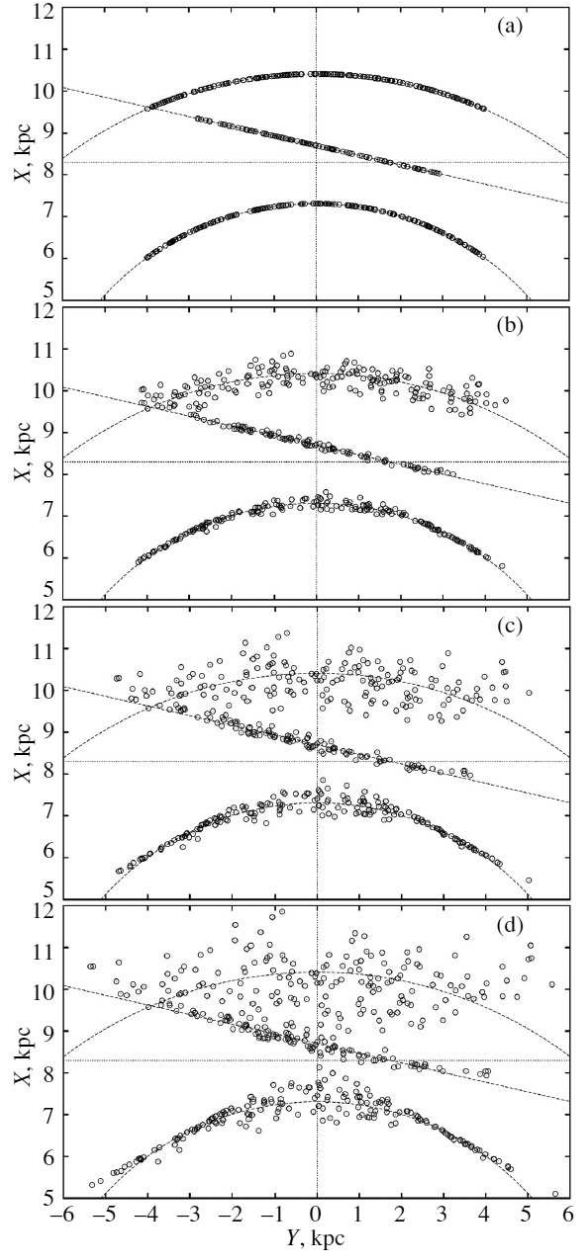}
 \caption{
Distributions of 280 model OSCs on the Galactic $XY$ plane (a)
obtained by the Monte Carlo method as a result of one realization
with the addition of random errors to the distances of 10\% (b),
20\% (c), and 30\% (d).
 }
  \label{f4}
 \end{center} }
 \end{figure}
%%%%%%%%%%%%%%%%%%%%%%%%%%%%%%%%%%%%%%%%%%%%%%%%%%%%%%%%%%%%%%%%%%%
%%%%%%%%%%%%%%%%%%%%%%%%%%%%%%%%%%%%%%%%%%%%%%%%%%%%%%%%%%%%%%%%%%%
 \begin{figure}
 {\begin{center}
 \includegraphics[width=0.5\textwidth]{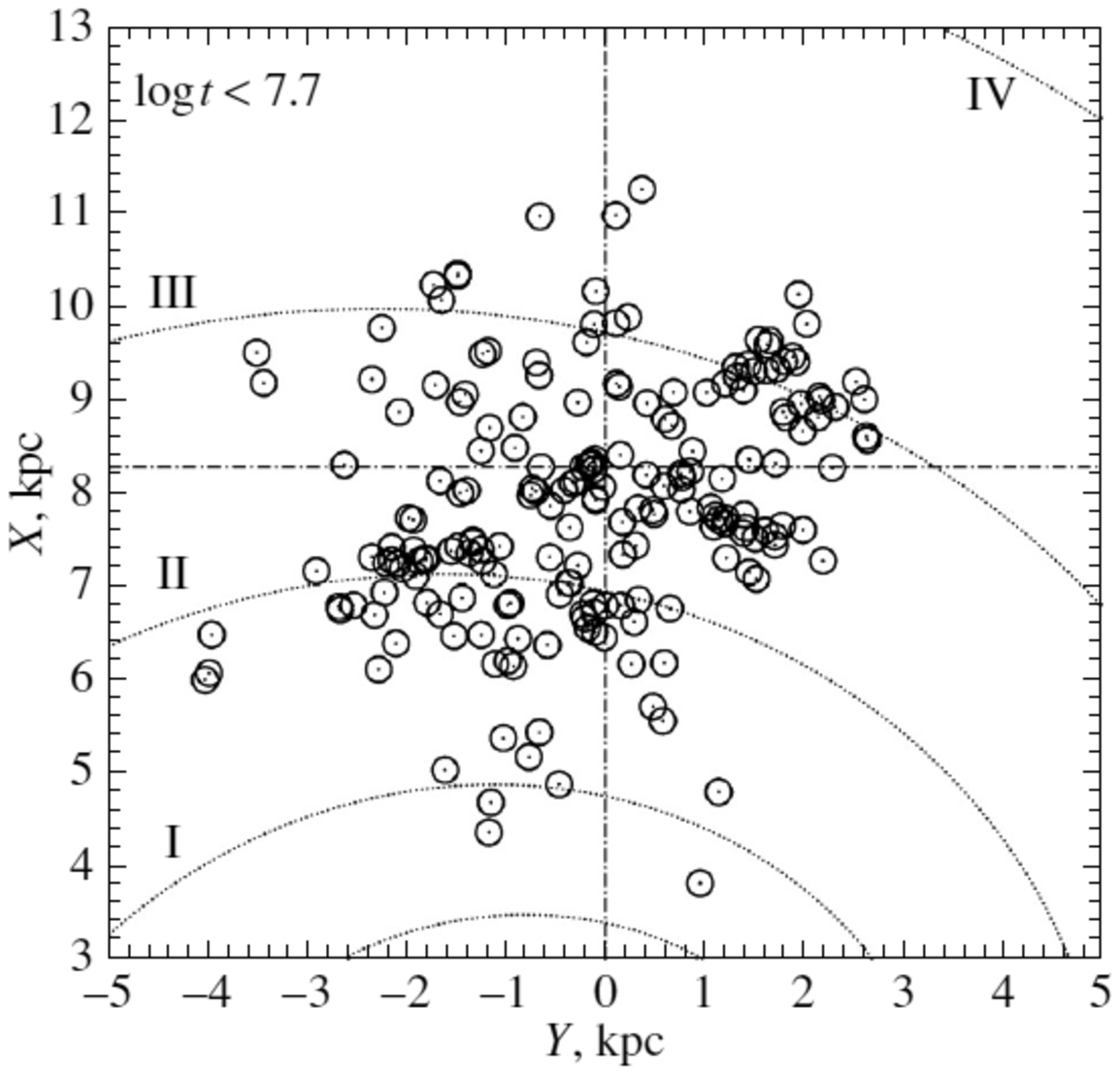}
 \caption{
Distributions of the sample of 209 young OSCs with known space
velocities on the Galactic $XY$ plane. The Sun's coordinates are
$(X,Y)=(8.3,0)$~kpc. The four-armed spiral pattern with the pitch
angle $i=-13^\circ$ found from masers (Bobylev and Bajkova 2014c)
is shown. The spiral arm segments are numbered by Roman numerals.
 }
  \label{f5}
 \end{center} }
 \end{figure}
%%%%%%%%%%%%%%%%%%%%%%%%%%%%%%%%%%%%%%%%%%%%%%%%%%%%%%%%%%%%%%%%%%%

 \subsection*{On the Geometry of the Spiral Pattern}
As can be seen from Fig. 1, the spiral pattern in the distribution
of OSCs is visible, but only in the sample of youngest clusters
$(\log t<7.7).$ However, there is significant noise even in the
distribution of these young OSCs.

We simulated the distribution of young clusters on the Galactic
$XY$ plane as a function of errors in their distances. For this
purpose we generated a model sample of 280 points corresponding to
the clusters populating the Local Arm (100 points) and two more
distant segments of the spiral arms, Perseus and
Carina--Sagittarius (90 points in each). The Local Arm was
represented by a line segment with a total length of 6 kpc that
was slightly displaced from the Sun toward the Galactic anticenter
and oriented at an angle of $13^\circ$ to the $Y$ axis (Bobylev
and Bajkova 2014d). In this model the segments of the Perseus
($X\approx9.5$~kpc) and Carina--Sagittarius ($X\approx7$ kpc)
spiral arms are represented as arcs of circumferences.

We generated three realizations by the Monte Carlo method with the
addition of random errors to the distances of 10\%, 20\%, and
30\%. The simulation results are presented in Fig. 4, where the
upper panel shows the initial model.

Our simulations showed the following: (i) if the objects lie on
the line of sight, as is the case for the edge segments of the
Carina.Sagittarius arm (II, $X\approx$7 kpc) and the Local Arm,
then the belonging to the specified model is retained even at
large random errors in the distances; (ii) from the distribution
of objects in the Perseus arm (III, $X\approx$9.5 kpc) that lie on
different lines of sight we can draw the conclusion about a
fundamental similarity to the distribution of the youngest $(\log
t<7.7)$ OSCs (Fig. 1) for a random error level of at least 20\%.

Figure 5 presents the distribution of the sample of 209 OSCs with
known space velocities on the Galactic $XY$ plane. In comparison
with the distribution of the youngest OSCs (Fig. 1), the
connection of OSCs with the plotted spiral pattern is seen more
clearly here. The Local Arm and part of the Carina--Sagittarius
arm are especially prominent. This is probably because the
selection of cluster members and the distance determination are
made more reliably at known line-of-sight velocities and proper
motions.

In Fig. 1 we have more cases where the candidates for clusters
were selected based on the proper motions of individual stars with
a comparatively low accuracy. This causes the errors in the
distance estimates to increase and, consequently, the spiral
pattern to be blurred. The motion of such clusters more likely
reflects the kinematics of the Galactic ``background'' than the
clusters themselves. However, such a smoothing effect is more
likely useful than harmful in determining the Galactic rotation
parameters.

The cluster age is also important. For example, Dias and L\'epine
(2005) showed that OSCs with an age of no more than 20 Myr are
closely connected with the spiral structure.

 \section*{DISCUSSION}
At present, there is a sample of more than 120 maser sources whose
trigonometric parallaxes were measured by the VLBI technique with
a high accuracy, with a mean error of $\pm$20 mas and, some of
them, with a record error of $\pm$5 mas. Honma et al. (2012)
estimated the Sun's velocity $V_0=238\pm14$ km s$^{-1}$ (for the
derived $R_0=8.05\pm0.45$~kpc) from their analysis of masers, Reid
et al. (2014) determined $V_0=240\pm8$ km s$^{-1}$ (for the
derived $R_0=8.34\pm0.16$ kpc), and, finally, Bobylev and Bajkova
(2014a) calculated $V_0=241\pm7$~km s$^{-1}$ (for the adopted
$R_0=8.3\pm0.2$~kpc). It can be seen that the value of this
velocity $V_0=236\pm6$ km s$^{-1}$ (for $R_0=8.3\pm0.2$ kpc) found
from a large amount of data on OSCs in solution (15) is in good
agreement with the present-day estimates obtained in other works.

We detected the influence of the spiral structure only in the
spatial distribution and kinematics of the youngest clusters whose
age does not exceed 50 Myr. It is interesting to compare the
parameters found from these clusters, such as the velocity
perturbation amplitudes $f_\theta,$ and $f_R,$ the wavelengths
$\lambda_\theta,$ and $\lambda_R,$ and the Sun's phases in the
spiral density wave  $(\chi_\odot)_\theta,(\chi_\odot)_R,$ with
those determined from other samples.

For example, Bobylev and Bajkova (2015a) analyzed spectroscopic
binaries and OB3 stars with the calcium distance scale based on a
spectral analysis. Based on a sample of spectroscopic binaries, we
found the following parameters for the model of a four-armed
spiral pattern $(m=4, R_0=8$ kpc):
 $f_R=9.5\pm1.5$ km s$^{-1}$ and
 $f_\theta=3.2\pm1.4$ km s$^{-1}$,
  $\lambda_R=2.8\pm0.5$ kpc $(i_R=-13^\circ\pm4^\circ)$ at the Sun's phase in the spiral
density wave $(\chi_\odot)_R=-95^\circ\pm15^\circ,$
$\lambda_\theta=2.6\pm0.4$ kpc $(i_\theta=-12^\circ\pm3^\circ)$ at
$(\chi_\odot)_\theta=-93^\circ\pm12^\circ.$ We obtained the
following estimates from a sample of OB3 stars with the calcium
distance scale: $f_R=11.8\pm1.3$ km s$^{-1}$ and
$\lambda_R=2.1\pm0.3$ kpc $(i_R=-9.5^\circ\pm1.7^\circ)$ at the
Sun's phase in the spiral density wave
$(\chi_\odot)_R=-86^\circ\pm7^\circ.$

Zabolotskikh et al. (2002) obtained $f_\theta=0.4\pm2.3$~km
s$^{-1}$, $f_R=6.6\pm2.5$~km s$^{-1}$, $i=-6.6^\circ\pm0.9^\circ$
and $\chi_\odot=-97^\circ\pm18^\circ$ $(m=2, R_0=7.5$~kpc) from a
sample of blue supergiants and
 $f_\theta=0.2\pm1.6$~km s$^{-1}$,
 $f_R=5.5\pm2.3$~km s$^{-1}$,
 $i=-12.2^\circ\pm0.7^\circ$ and $\chi_\odot=-88^\circ\pm14^\circ$
 $(m=4, R_0=7.5$ kpc) from a sample of open clusters. Note that
Zabolotskikh et al. (2002) determined the parameters of the spiral
density wave simultaneously with the Galactic rotation parameters,
i.e., no spectral analysis was used. Dambis et al. (2015) found
the phase  $\chi_\odot=-121^\circ\pm3^\circ$ $(m=4, R_0=7.1$~kpc)
by analyzing the currently most complete kinematic sample of
classical Cepheids. Using a sample of 107 masers with measured
trigonometric parallaxes and based on a spectral analysis, Bobylev
and Bajkova (2015b) estimated
 $f_\theta=6.0\pm2.6$~km s$^{-1}$ and $f_R=7.2\pm2.2$~km s$^{-1}$,
 $\lambda_\theta=3.2\pm0.5$~kpc and $\lambda_R=3.0\pm0.6$~kpc $(m=4, R_0=8.0$~kpc),
 $(\chi_\odot)_\theta=-79^\circ\pm14^\circ$ and $(\chi_\odot)_R=-199^\circ\pm16^\circ$.

We can see that the wavelengths $\lambda_\theta,$ $\lambda_R$ and
the Sun’s phases in the spiral density wave $(\chi_\odot)_\theta,$
$(\chi_\odot)_R$ found in this paper from clusters are in good
agreement with the results of the analysis of OB stars, Cepheids,
and masers. The fairly large amplitude of the perturbations in the
tangential velocities of OSCs $f_\theta=5.6\pm1.6$~km s$^{-1}$,
which from samples of OB stars usually turns out to be no
significantly different from zero, is of considerable interest.

The connection of the spatial distribution of young OSCs with the
spiral structure was pointed out by many authors (Dias and
L\'epine 2005; Naoz and Shaviv 2007; Loktin and Popova 2007; Griv
et al. 2014). Since the radius of the sample with reliable OSC
distance estimates is small (approximately 3.5 kpc), the estimates
of the parameters of the spiral structure based on OSCs are
unreliable. For example, from their analysis of OSCs Popova and
Loktin (2005) found the pitch angle $i=-21.5^\circ$, while based
on a joint analysis of OSCs, HII clouds, and Cepheids these
authors hypothesized the existence of a 12-armed spiral structure
in the Galaxy (Loktin and Popova 2007).

Hou and Han (2014) gave an overview of the present-day estimates
for the parameters of the spiral structure obtained from objects
(HI clouds, HII regions, molecular clouds, and masers) distributed
over the entire Galaxy. These authors justify the four-armed model
with a pitch angle close to $-13^\circ.$ Based on a sample of 565
classical Cepheids, Dambis et al. (2015) found the pitch angle
$i=-9.5^\circ\pm0.1^\circ$ within the four-armed model of the
spiral structure. According to the latest estimate by Valle\'e
(2015) obtained by analyzing various indicators, the pitch angle
of the four-armed spiral pattern in the Galaxy is
$i=-13.1^\circ\pm0.6^\circ.$

 \section*{CONCLUSIONS}
The Galactic rotation parameters were redetermined using a large
sample of open star clusters from the MWSC (Milky Way Star
Clusters) catalogue produced by Kharchenko et al. (2013). An
important advantage of the catalogue is the homogeneity of its
proper motions that were obtained using the PPMXL catalogue
(R\"oser et al. 2010).

The circular rotation velocity of the solar neighborhood around
the Galactic center was found from data on $\sim$2000 OSCs to be
$V_0=236\pm6$~km s$^{-1}$ for the adopted Galactocentric distance
of the Sun $R_0=8.3\pm0.2$ kpc. The Oort constants
$A=-14.52\pm0.32$~km s$^{-1}$ kpc$^{-1}$ and $B=13.95\pm0.48$ km
s$^{-1}$ kpc$^{-1}$ found in this solution show that the Galactic
rotation curve is nearly flat in the solar neighborhood. This
solution was obtained using almost all clusters from the
catalogue, because they are all provided with the proper motions.
In addition, the line-of-sight velocities of the clusters were
also involved in the solution.

Remarkably, the Galactic rotation parameters ($\Omega_0,$
$\Omega'_0,$ $\Omega''_0,$ $V_0$, $A,$ $B$) that were determined
from samples of clusters with various ages are very stable. In
addition, the values of $\Omega_0,$ $V_0$, $A,$ and $B$ are close
to those found by various authors from masers with measured
trigonometric parallaxes. Obviously, this favorably characterizes
the system of proper motions of the PPMXL catalogue.

The parameters of the Galactic spiral density wave satisfying the
linear Lin.Shu model were found from the series of residual
tangential, $\Delta V_{circ},$ and radial, $V_R,$ velocities for
the sample of youngest clusters $(\log t<7.7)$ using a periodogram
analysis. The amplitudes of the tangential and radial velocity
perturbations are
 $f_\theta=5.6\pm1.6$~km s$^{-1}$ and
 $f_R=7.7\pm1.4$~km s$^{-1}$, respectively; the perturbation
wavelengths are $\lambda_\theta=2.6\pm0.5$~kpc
($i_\theta=-11^\circ\pm2^\circ$) and $\lambda_R=2.1\pm0.5$~kpc
($i_R=-9^\circ\pm2^\circ$) for the adopted four-armed model of the
spiral pattern $(m=4).$ The Sun's phase in the spiral density wave
is $(\chi_\odot)_\theta=-62^\circ\pm9^\circ$ and
$(\chi_\odot)_R=-85^\circ\pm10^\circ$ from the residual tangential
and radial velocities, respectively. No influence of the spiral
structure in the velocities and positions of older clusters $(\log
t>7.7)$ was detected.

Monte Carlo simulations of the distribution of clusters in space
showed the errors in the distances to be, on average, no less than
20\%.

 \subsection*{ACKNOWLEDGMENTS}
We are grateful to the referees for their helpful remarks that
contributed to an improvement of this paper. This work was
supported by the ``Transitional and Explosive Processes in
Astrophysics'' Program P--41 of the Presidium of Russian Academy
of Sciences.

 \bigskip{REFERENCES}\medskip
 {\small

1. L.H. Amaral and J.R.D. L\'epine, Mon. Not. R. Astron. Soc. 286,
885 (1997).

2. A.T. Bajkova and V.V. Bobylev, Astron. Lett. 38, 549 (2012).

3. A.T. Bajkova and V.V. Bobylev, Baltic Astron. 24, 43 (2015).

4. V.V. Bobylev, A.T. Bajkova, and S.V. Lebedeva, Astron. Lett.
33, 720 (2007).

5. V.V. Bobylev, A.T. Bajkova, and A.S. Stepanishchev, Astron.
Lett. 34, 515 (2008).

6. V.V. Bobylev and A.T. Bajkova, Astron. Lett. 38, 638 (2012).

7. V.V. Bobylev and A.T. Bajkova, Astron. Lett. 39, 532 (2013).

8. V.V. Bobylev and A.T. Bajkova, Astron. Lett. 40, 389 (2014a).

9. V.V. Bobylev and A.T. Bajkova, Astron. Lett. 40, 773 (2014b).

10. V.V. Bobylev and A.T. Bajkova, Mon. Not. R. Astron. Soc. 437,
1549 (2014c).

11. V.V. Bobylev and A.T. Bajkova, Astron. Lett. 40, 783 (2014c).

12. V.V. Bobylev and A.T. Bajkova, Astron. Lett. 41, 473 (2015a).

13. V.V. Bobylev and A.T. Bajkova, Mon. Not. R. Astron. Soc. 447,
L50 (2015b).

14. C. Bonatto, L.O. Kerber, E. Bica, and B.X. Santiago, Astron.
Astrophys. 446, 121 (2006).

15. J. Borissova, A.-N. Chen\'e, S. Ramirez Alegria, S. Sharma,
J.R.A. Clarke, R. Kurtev, I. Negueruela, A. Marco, P. Amigo, et
al., Astron. Astrophys. 569, A24 (2014).

16. D. Camargo, C. Bonatto, and E. Bica, Mon. Not. R. Astron. Soc.
450, 4150 (2015).

17. D. Camargo, E. Bica, and C. Bonatto, Mon. Not. R. Astron. Soc.
455, 3126 (2016).

18. B. Chaboyer, E.M. Green, and J. Liebert, Astron. J. 117, 1360
(1999).

19. L. Chen, J.L. Hou, and J.J. Wang, Astron. J. 125, 1397 (2003).

20. G.C. Clayton and E.L. Fitzpatrick, Astron. J. 93, 157 (1987).

21. N.J.G. Cross, R.S. Collins, R.G. Mann, M.A. Read, E.T.W.
Sutorius, R.P. Blake, M. Holliman, N.C. Hambly, et al., Astron.
Astrophys. 548, A119 (2012).

22. A.K. Dambis, L.N. Berdnikov, Yu.N. Efremov, A.Yu. Knyazev,
A.S. Rastorguev, E.V. Glushkova, V.V. Kravtsov, D.G. Terner, D.D.
Madzhess, and R. Sefako, Astron. Lett. 41, 489 (2015).

23. W.S. Dias, B.S. Alessi, A. Moitinho, and J.R.D. L\'epine,
Astron. Astrophys. 389, 871 (2002).

24. W.S. Dias and J.R.D. L\'epine, Astrophys. J. 629, 825 (2005).

25. P. Fedorov, A. Myznikov, and V. Akhmetov, Mon. Not. R. Astron.
Soc. 393, 133 (2009).

26. P. Fedorov, V. Akhmetov, and V.V. Bobylev, Mon. Not. R.
Astron. Soc. 416, 403 (2011).

27. E.D. Friel, Ann. Rev. Astron. Astrophys. 33, 381 (1995).

28. D. Froebrich, A. Scholz, and C.L. Raftery, Mon. Not. R.
Astron. Soc. 374, 399 (2007).

29. S. Gillessen, F. Eisenhauer, T.K. Fritz, H. Bartko, K.
Dodds-Eden, O. Pfuhl, T. Ott, and R. Genzel, Astroph. J. 707, L114
(2009).

30. E.V. Glushkova, A.K. Dambis, A.M. Mel’nik, and A.S.
Rastorguev, Astron. Astrophys. 329, 514 (1998).

31. E.V. Glushkova, S.E. Koposov, I.Yu. Zolotukhin, Yu.V.
Beletskii, A.D. Vlasov, and S.I. Leonova, Astron. Lett. 36, 75
(2010).

32. E. Griv, C.-C. Lin, C.-C. Ngeow, and I.G. Jiang, New Astron.
29, 9 (2014).

33. The HIPPARCOS and Tycho Catalogues, ESA SP--1200 (1997).

34. M. Honma, T. Nagayama, K. Ando, T. Bushimata, Y.K. Choi, T.
Handa, T. Hirota, H. Imai, T. Jike, et al., Publ. Astron. Soc.
Jpn. 64, 136 (2012).

35. L.G. Hou and J.L. Han, Astron. Astrophys. 569, A125 (2014).

36. Y.C. Joshi, Mon. Not. R. Astron. Soc. 362, 1259 (2005).

37. Y.C. Joshi, Mon. Not. R. Astron. Soc. 378, 768 (2007).

38. T.C. Junqueira, C. Chiappini, J.R.D. L\'epine, I. Minchev, and
B.X. Santiago, Mon. Not. R. Astron. Soc. 449, 2336 (2015).

39. N.V. Kharchenko and S. R\"oser, Kinem. Phys. Celest. Bodies
17, 409 (2001).

40. N.V. Kharchenko, A.E. Piskunov, S. R\"oser, E. Schilbach, and
R.-D. Scholz, Astron. Astrophys. 438, 1163 (2005a).

41. N.V. Kharchenko, A.E. Piskunov, S. R\"oser, E. Schilbach, and
R.-D. Scholz, Astron. Astrophys. 440, 403 (2005b).

42. N.V. Kharchenko, R.-D. Scholz, A.E. Piskunov, S. R\"oser, and
E. Schilbach, Astron. Nachr. 328, 889 (2007).

43. N.V. Kharchenko, A.E. Piskunov, S. Roeser, E. Schilbach, and
R.-D. Scholz, Astron. Astrophys. 558, 53 (2013).

44. S. Koposov, E. Glushkova, and I. Zolotukhin, Astron. Nachr.
326, 597 (2005).

45. S.E. Koposov, E.V. Glushkova, and I.Yu. Zolotukhin, Astron.
Astrophys. 486, 771 (2008).

46. J.R.D. L\'epine, W.S. Dias, and Yu. Mishurov, Mon. Not. R.
Astron. Soc. 386, 2081 (2008).

47. C.C. Lin and F.H. Shu, Astrophys. J. 140, 646 (1964).

48. A.V. Loktin and G.V. Beshenov, Astron. Rep. 47, 6 (2003).

49. A.V. Loktin and M.E. Popova, Astron. Rep. 51, 364 (2007).

50. A. Maeder and J.C. Mermilliod, Astron. Astrophys. 93, 136
(1981).

51. P.G.P. Moroni and O. Straniero, Astrophys. J. 581, 585 (2002).

52. S. Naoz and N.J. Shaviv, New Astron. 12, 410 (2007).

53. M. Netopil, E. Paunzen, U. Heiter, and C. Soubiran, Astron.
Astrophys. 585, 150 (2016).

54. R.L. Phelps, Astrophys. J. 483, 826 (1997).

55. A.E. Piskunov, N.V. Kharchenko, S. R\"oser, E. Schilbach, and
R.-D. Scholz, Astron. Astrophys. 445, 545 (2006).

56. M.E. Popova and A.V. Loktin, Astron. Lett. 31, 171 (2005).

57. M.J. Reid, K.M. Menten, A. Brunthaler, X.W. Zheng, T.M. Dame,
Y. Xu, Y. Wu, B. Zhang, et al., Astrophys. J. 783, 130 (2014).

58. S. R\"oser, M. Demleitner, and E. Schilbach, Astron. J. 139,
2440 (2010).

59. S. Schmeja, N.V. Kharchenko, A.E. Piskunov, S. R\"oser, E.
Schilbach, D. Froebrich, and R.-D. Scholz, Astron. Astrophys. 568,
A51 (2014).

60. R.-D. Scholz, N.V. Kharchenko, A.E. Piskunov, S.R\"oser, and
E. Schilbach, Astron. Astrophys. 581, A39 (2015).

61. M.F. Skrutskie, R. M. Cutri, R. Stiening, M.D. Weinberg, S.
Schneider, J.M. Carpenter, C. Beichman, R. Capps, et al., Astron.
J. 131, 1163 (2006).

62. R.J. Trumpler, Publ. Astron. Soc. Pacif. 42, 214 (1930).

63. B.A. Twarog and B.J. Anthony-Twarog, Astron. J. 97, 759
(1989).

64. J.P. Vall\'ee, Mon. Not. R. Astron. Soc. 450, 4277 (2015).

65. E.L. Wright, P.R.M. Eisenhardt, A.K. Mainzer, M.E. Ressler,
R.M. Cutri, T. Jarrett, J.D. Kirkpatrick, D. Padgett, et al.,
Astron. J. 140, 1868 (2010).

66. M.V. Zabolotskikh, A.S. Rastorguev, and A.K. Dambis, Astron.
Lett. 28, 454 (2002).
 }

\end{document}